\documentclass{Interspeech2024}
\usepackage{algorithmic}
\usepackage{algorithm}
\usepackage{multirow}
\usepackage{cite}
\usepackage[dvipsnames]{xcolor}
\usepackage{tablefootnote}

\interspeechcameraready

\title{Contextualized End-to-end Automatic Speech Recognition with \\ Intermediate Biasing Loss}

\name[affiliation={1}]{Muhammad}{Shakeel}
\name[affiliation={1}]{Yui}{Sudo}
\name[affiliation={2}]{Yifan}{Peng}
\name[affiliation={2}]{Shinji}{Watanabe}

\address{
  $^1$Honda Research Institute Japan Co., Ltd., Saitama, Japan\\
  $^2$Carnegie Mellon University, Pittsburgh, USA}
\email{ \{shakeel.muhammad,yui.sudo\}@jp.honda-ri.com, \{yifanpen,swatanab\}@andrew.cmu.edu}


\begin{document}

\maketitle

\begin{abstract}
Contextualized end-to-end automatic speech recognition has been an active research area, with recent efforts focusing on the implicit learning of contextual phrases based on the final loss objective. However, these approaches ignore the useful contextual knowledge encoded in the intermediate layers. We hypothesize that employing explicit biasing loss as an auxiliary task in the encoder intermediate layers may better align text tokens or audio frames with the desired objectives. Our proposed intermediate biasing loss brings more regularization and contextualization to the network. Our method outperforms a conventional contextual biasing baseline on the LibriSpeech corpus, achieving a relative improvement of 22.5\% in biased word error rate (B-WER) and up to 44\% compared to the non-contextual baseline with a biasing list size of 100. Moreover, employing RNN-transducer-driven joint decoding further reduces the unbiased word error rate (U-WER), resulting in a more robust network.
\end{abstract}

\section{Introduction}
\label{sec:intro}
End-to-end (E2E) automatic speech recognition (ASR) \cite{asr_review1,asr_review2} has been a focus of study for its effectiveness in transcribing speech, resulting in substantial reductions in word error rates (WERs). These improvements are made possible by state-of-the-art architectures, including connectionist temporal classification (CTC) \cite{ctc1}, recurrent neural network transducers (RNN-transducer) \cite{rnnt1,rnnt2}, and attention-based \cite{attention1} models. Hybrid methods \cite{ctc-attention1,4D} that combine these techniques, occasionally supplemented with auxiliary learning \cite{interctc1,interctc2,interctc3} such as intermediate CTC (InterCTC) loss \cite{interctc4}, has further boosted performance. However, the effective integration of contextual information, such as specialized vocabulary (referred to as a biasing list), remains an open challenge and often results in out-of-context transcriptions. Thus, it is crucial for E2E ASR architectures to effectively incorporate context-specific prior knowledge to improve transcription accuracy in specific domains.

Several studies have addressed the challenge of incorporating contextual information by investigating modifications to existing architectures, leading to two broad categories of approaches: 1) Shallow fusion-based methods involve transforming bias phrases into weighted finite-state transducers (WFSTs) at word or subword boundaries \cite{shallow1,shallow2,shallow3}, creating a contextual language model (LM) for rescoring. Alternatively, some methods \cite{shallow4,shallow5,shallow6} model the bias phrases during training, inserting user-defined bias phrases as WFST graphs at appropriate positions in the beam search. These methods either utilize separately trained user-dependent external language models (LMs) or bias phrase information to enhance the performance of user-defined bias phrases through on-the-fly rescoring of the initial ASR hypothesis. Despite their effectiveness, these methods have inherent limitations, such as potential under- or over-biasing due to a lack of joint optimization during training, and may require heuristics to adjust the weights of the contextual LMs. 2) In contrast, neural deep-biasing methods \cite{bias1,bias2,bias3,bias3a,bias4,bias4a} directly or indirectly integrate bias phrases with audio features in the intermediate layers \cite{bias13,bias14}, often employing attention mechanisms \cite{bias5,bias6,bias7,bias7a,bias8,bias9,bias12}. However, they require additional bias phrase information \cite{shallow7a,shallow7b,bias10,bias10a,bias10b,bias11} either to the encoder or decoder during training or inference. In an alternative approach, auxiliary loss is applied either to the \textit{last} layer of the encoder \cite{bias9} or decoder \cite{bias12} to explicitly supervise bias phrases. This improves contextualization by mapping higher-level abstract representations of the audio input to bias phrases. However, it may limit the model's ability to explicitly align bias phrases with the input audio, relying on bias phrase prediction from learned higher-level abstract representations while ignoring lower levels during training. To bridge this gap between input and output representations, and thus improve contextualization, an E2E ASR model should be trained to explicitly align bias phrases within the \textit{intermediate} levels of audio abstraction. This would enable more effective extraction and learning of contextual representations.

In this work, we present a novel method to improve contextualization in E2E ASR models. Our approach proposes to involve an auxiliary task that directly aligns bias phrases with the intermediate audio representations and can be adopted to any E2E ASR architecture. Our key contributions are:
\begin{itemize}
    \item We introduce a new auxiliary task, the intermediate biasing (IB) loss based on CTC, allowing for improved contextualization by utilizing information across the  intermediate representations of the audio encoder.
    \item We show that the IB loss can be adapted to various architectures such as CTC, attention, and RNN-transducer. In this paper, we specifically introduce its application to CTC \cite{interctc4} and contextual RNN-transducer-based \cite{bias4} models.
    \item We achieved substantial reductions in word/character error rates (WER/CER) and biased word/character error rates (B-WER/B-CER) on the Librispeech and in-house datasets compared to conventional approaches. Additionally, RNN-transducer-driven joint decoding with CTC is introduced that further minimizes the unbiased word/character error rates (U-WER/U-CER) to mitigate the effect of biasing on unbiased words/characters when the biasing list size is large.
\end{itemize}
\begin{figure*}[t]
\begin{center}
{\resizebox*{13.5cm}{!}{\includegraphics[width=\textwidth]{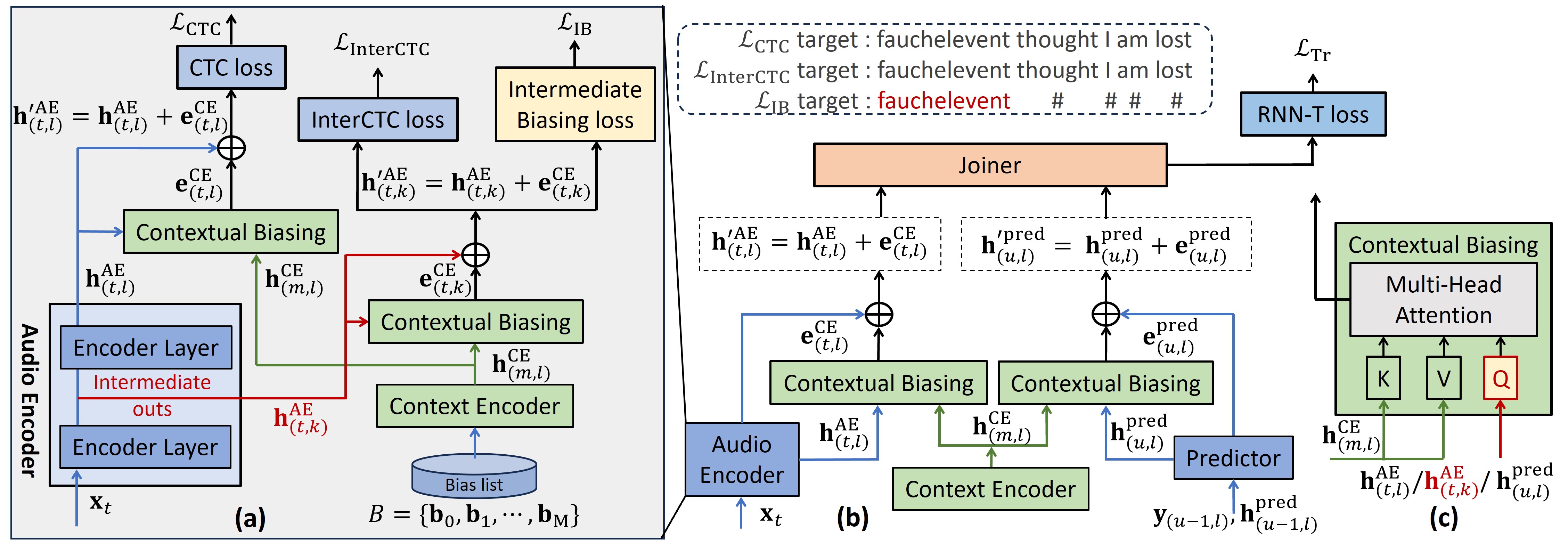}}}
\vspace*{-4mm}
\caption{Proposed model: (a) Audio encoder with the proposed intermediate biasing (IB) loss, (b) integrated with the InterCTC-based \cite{interctc4} contextual RNN-transducer\cite{bias4}, (c) employing a multi-head attention-based (MHA) contextual biasing (CB) module.}
\label{fig:biasloss} 
\end{center}
\vspace*{-9mm}
\end{figure*}
\vspace*{-2mm}
\section{Preliminary}
\vspace*{-1mm}
We present an E2E ASR architecture consisting of an audio encoder and an InterCTC \cite{interctc4} loss within both CTC and RNN-transducer \cite{rnnt1} formulations.
\vspace*{-3mm}
\subsection{Audio Encoder}
\label{sec:audioencoder}
\vspace*{-1mm}
The audio encoder \begin{math} \mathrm{AudioEnc(\cdot)}\end{math} comprises $l$ encoder layers, where $l$ is a layer index from 1 to $N$. We leverage the conformer \cite{conformer} architecture with Transformer and convolutional layers operating on a $T$-length input acoustic feature sequence \begin{math} X = \{\mathbf{x}_{t} \in \mathbb{R}^{D}|t = 1,\cdots,T\} \end{math}, and transforms a subsampled $T'(<T)$-length feature sequence to a high-level representation \begin{math} H_{(N)}^{\text{AE}} = \{\mathbf{h}_{(t,l)}^{\text{AE}} \in \mathbb{R}^{S}|t = 1,\cdots,T'\} \end{math}, given by:
\vspace*{-2mm}
\begin{equation}
\mathbf{h}_{(t,l)}^{\text{AE}} = \mathrm{AudioEnc}(\mathbf{x}_{t}),
\label{eq:audioencoder}
\vspace*{-2mm}
\end{equation}
where $\mathbf{x}_{t}$ and $\mathbf{h}_{(t,l)}^{\text{AE}}$ are a $D$ and $S$-dimensional acoustic features at frame $t$ for layer $l$, respectively.

Given an input sequence and its corresponding $U$-length output sequence \begin{math} Y = \{y_{(u,l)} \in \mathcal{V}|u = 1,\cdots,U\} \end{math}, where $y_{(u,l)}$ is an output token at position $u$ in the vocabulary $\mathcal{V}$ for layer $l$, the audio encoder is trained using the CTC loss ($\mathcal{L}_{\text{CTC}}$) objective at the $last$ ($N$) layer. The training process is optimized by minimizing the negative log-likelihood given by:
\vspace*{-2mm}
\begin{equation}
\mathcal{L}_{\text{CTC}} = - \log \mathrm{P}_{\text{CTC}}(y_{(u,l)}|\mathbf{h}_{(t,l)}^{\text{AE}}).
\label{eq:ctc_off_loss}
\vspace*{-2mm}
\end{equation}
\subsection{InterCTC}
\label{sec:interctc}
\vspace*{-1mm}
The InterCTC loss employs auxiliary CTC losses to the intermediate hidden states of the audio encoder, in addition to the primary CTC loss applied at the $last$ ($N$) layer, as described in Eq. \eqref{eq:ctc_off_loss}. This regularizes the model parameters and is given by: 
\vspace*{-2mm}
\begin{equation}
\mathcal{L}_{\text{InterCTC}} = \dfrac{1}{\mathcal{K}} \sum_{k \in \mathcal{K}} - \log \mathrm{P}_{\text{CTC}}(y_{(u,k)}|\mathbf{h}_{(t,k)}^{\text{AE}}),
\label{eq:interctc_loss}
\vspace*{-2mm}
\end{equation}
where $\mathcal{K}$ $\subseteq$ $\{1,\cdots,N-1\}$ represents the positions of the intermediate layers where InterCTC loss is applied.  
We optimize the audio encoder loss ($\mathcal{L}_{\text{AE}}$) by combining Eqs. \eqref{eq:ctc_off_loss} and \eqref{eq:interctc_loss} using a tunable hyper-parameter $\lambda_{\text{ic}}$:
\vspace*{-2mm}
\begin{equation}
\mathcal{L}_\text{AE} = (1 - \lambda_{\text{ic}}) \mathcal{L}_{\text{CTC}} + \lambda_{\text{ic}} \mathcal{L}_{\text{InterCTC}}.
\label{eq:ae_interctc_loss}
\vspace*{-3mm}
\end{equation}
\subsection{RNN-transducer}
\label{sec:transducer}
\vspace*{-1mm}
A RNN-transducer \cite{rnnt1} typically consists of an audio encoder, a predictor network, and a joiner network. The audio encoder outputs a hidden state vector \begin{math} \mathbf{h}_{(t,l)}^{\text{AE}} \end{math} for layer $l$ at frame $t$ as described in Section \ref{sec:audioencoder}. The predictor network, denoted as \begin{math} \mathrm{Predictor(\cdot)} \end{math}, takes the previous non-blank label $y_{(u-1,l)}$ as input in Eq. \eqref{eq:predictor} and generates a vector representation \begin{math} \mathbf{h}_{(u,l)}^{\text{pred}} \end{math} at position $u$. The joiner network, denoted as \begin{math} \mathrm{Joiner(\cdot)} \end{math}, combines the output of the two networks using a linear layer and predicts the output token in the form of posteriors using Eq. \eqref{eq:jointnetwork}: 
\vspace*{-1mm}
\begin{equation}
\mathbf{h}_{(u,l)}^{\text{pred}} = \mathrm{Predictor}({y}_{{(u-1,l)}},\mathbf{h}_{{(u-1,l)}}^{\text{pred}}),
\label{eq:predictor}
\vspace*{-2mm}
\end{equation}
\begin{equation}
\mathrm{P}(y_{(u,l)}|\mathbf{x}_{t},y_{(1:u-1,l)}) = \mathrm{Softmax}(\mathrm{Joiner}(\mathbf{h}_{(t,l)}^{\text{AE}},\mathbf{h}_{(u,l)}^{\text{pred}})),
\label{eq:jointnetwork}
\end{equation}
The RNN-transducer loss ($\mathcal{L}_{\text{Tr}}$) then updates the model parameters by minimizing the negative log-likelihood given by:
\vspace*{-1mm}
\begin{equation}
\mathcal{L}_{\text{Tr}} = - \log \mathrm{P}_{\text{Tr}}(y_{(u,l)}|\mathbf{h}_{(t,l)}^{\text{AE}}).
\label{eq:transducer_loss}
\vspace*{-1mm}
\end{equation}
\vspace*{-5mm}
\section{Proposed architecture}
\vspace*{-1mm}
Fig. \ref{fig:biasloss} provides an overview of the proposed model. In this section, we explain the IB loss, context encoder, and a contextual biasing (CB) module (Fig. \ref{fig:biasloss}(c)) within the audio encoder (Fig. \ref{fig:biasloss}(a)) and decoder components (Fig. \ref{fig:biasloss}(b)) for contextualization.
\vspace*{-6mm}
\subsection{Context Encoder}
\label{sec:context-encoder}
\vspace*{-1mm}
Given a list of $M$ bias phrases \begin{math} B = (\mathbf{b}_{0},\mathbf{b}_{1},\cdots, \mathbf{b}_{M})\end{math}, each bias phrase $\mathbf{b}_{m}$, where $m \in M$, is processed by the context encoder using a BiLSTM model \cite{bias4}. Here, $\mathbf{b}_{0}$ represents $\texttt{<no\_bias>}$ option of not using the context, which is similar to the $\texttt{<blank>}$ token for CTC. Moreover, each bias phrase $\mathbf{b}_{m} \in \mathbb{R}^{L_{\text{max}}}_\mathrm{+}$ consists of a token sequence of length $L_{\text{max}}$. These token sequences are obtained by zero-padding the extracted phrases to a fixed length $L_{\text{max}}$ greater than zero ($+$), during the training phase. The context encoder in Eq. \eqref{eq:bilstm} then encodes each phrase into a $S$-dimensional fixed-size phrase vector embedding $H^{\text{CE}} = \{\mathbf{h}^\text{CE}_{(m,l)} \in \mathbb{R}^{(M+1) \times S} |m = 1, \cdots,M\}$, given by:
\vspace*{-2mm}
\begin{equation}
\mathbf{h}^{\text{CE}}_{(m,l)} = \mathrm{BiLSTM}(\mathbf{b}_{(m,l)}) .
\label{eq:bilstm}
\end{equation}
The final hidden state of the BiLSTM is then fed to the CB module to obtain a context-aware hidden representation. Since bias list may not always be relevant, the introduction of the dummy $\texttt{<no\_bias>}$ token enables the model to learn when to enable or disable the biasing option during training and inference. 
\vspace*{-2mm}
\subsection{Contextual Biasing}
\label{sec:CB}
\vspace*{-1mm}
Given the intermediate representations of the audio encoder $\mathbf{h}_{(t,l)}^{\text{AE}}$ in Eq. \eqref{eq:audioencoder} and phrase embedding of the context encoder $\mathbf{h}_{(m,l)}^{\text{CE}}$ in Eq. \eqref{eq:bilstm}, we insert a CB module employing cross-attention mechanism (Fig. \ref{fig:biasloss}(a)) to compute the attention scores $\mathbf{A}_{(t,l)} \in \mathbb{R}^{T' \times (M+1)}$ in Eq. \eqref{eq:softmax} for layer $l$ at a given frame $t$:
\vspace*{-2mm}
\begin{equation}
\mathbf{A}_{(t,l)} = \mathrm{softmax} \left( \frac{\mathbf{h}_{(t,l)}^{\text{AE}}\mathbf{W}^{\text{Q}}(\mathbf{h}_{(m,l)}^{\text{CE}}\mathbf{W}^{\mathrm{K}})^\mathrm{T}}{ \sqrt{\mathrm{S}}}\right),
\label{eq:softmax}
\vspace*{-3mm}
\end{equation}
where $\mathbf{h}_{(t,l)}^{\text{AE}}$ serves as the attention query, $\mathbf{h}_{(m,l)}^{\text{CE}}$ acts as the keys and $\mathbf{W}^{\mathrm{Q}}$, $\mathbf{W}^{\mathrm{K}} \in \mathbb{R}^{S \times S}$ are the projection matrices that transform query and key. For simplicity, we define a single-head attention mechanism, however, this formulation extends to a multi-head attention (MHA) scenario.
\vspace*{-2mm}
\begin{equation}
\mathbf{e}_{(t,l)}^{\text{CE}} = \mathbf{A}_{(t,l)}\mathbf{h}_{(m,l)}^{\text{CE}}\mathbf{W}^\mathbf{V}; \quad
\mathbf{h}_{(t,l)}^{'\text{AE}} = \mathbf{h}_{(t,l)}^{\text{AE}} + \mathbf{e}_{(t,l)}^{\text{CE}}.
\label{eq:crossattention}
\end{equation}
We compute the biasing vector $\mathbf{e}_{(t,l)}^{\text{CE}}$ in Eq. \eqref{eq:crossattention} using a weighted sum of attention scores given the projection matrix $\mathbf{W}^\mathbf{V} \in \mathbb{R}^{S \times S} $ for value. We update the context-aware representation $\mathbf{h}_{(t,l)}^{'\text{AE}}$ (Fig. \ref{fig:biasloss}(c)) using the element wise addition between the hidden states of the audio encoder $ \mathbf{h}_{(t,l)}^{\text{AE}} $ and the context encoder $\mathbf{e}_{(t,l)}^{\text{CE}}$.
\subsection{Intermediate Biasing Loss}
\label{sec:ibl}
\vspace*{-1mm}
Prior studies \cite{bias9,bias12} show that using predicted posteriors and contextual phrases to compute CTC loss can improve contextual adaptation in E2E ASR models. However, their investigation is limited to the \textit{last} ($N$) layer of the encoder and shows minimal impact on contextual recognition when presented with a large list of bias phrases.  We address these limitations by adapting the loss function to the \textit{intermediate} layers of the encoder.

In this work, we employ the IB loss to introduce phrase-level contextual modeling to the ASR model. It allows us to extract bias embeddings from the utterance. For example, given a reference bias phrase $\texttt{"\textcolor{red}{fauchelevent}"}$ in the reference transcript $\texttt{"\textcolor{red}{fauchelevent}}$ $\texttt{thought}$ $\texttt{I am lost"}$, we generate a training target for the IB loss $\texttt{"\textcolor{red}{fauchelevent}}$ $\texttt{\#}$ $\texttt{\#}$ $\texttt{\#}$ $\texttt{\#"}$ by retaining the tokens of the target phrase and substituting the other tokens with a dummy token $\texttt{\#}$ (representing $\texttt{<no-bias>}$ output).  
We compute the IB loss based on the CTC objective in Eq. \eqref{eq:interbias_loss} for the bias phrases $\mathbf{b}_{(m,k)}$ (Section \ref{sec:context-encoder}). 
\vspace*{-3mm}
\begin{equation}
\mathcal{L}_{\text{IB}} = \dfrac{1}{\mathcal{K}} \sum_{k \in \mathcal{K}} - \log \mathrm{P}_{\text{ctc}}(\mathbf{b}_{(m,k)} | \mathbf{h}_{(t,k)}^{'\text{AE}}).
\label{eq:interbias_loss}
\vspace*{-2mm}
\end{equation}
We optimize the IB loss at a subset of layers $\mathcal{K}$ (Section \ref{sec:interctc}) by obtaining the intermediate context-aware representations $\mathbf{h}_{(t,k)}^{'\text{AE}}$ processed via CB module using Eq. \eqref{eq:crossattention} for explicit alignment. We optimize the audio encoder loss using Eqs.\eqref{eq:ae_interctc_loss} and \eqref{eq:interbias_loss} with tunable hyper-parameters $\lambda_{\text{ic}}$ and $\lambda_{\text{ib}}$:
\vspace*{-2mm}
\begin{equation}
\vspace*{-1mm}
\mathcal{L}_\text{AE} = (1 - \lambda_{\text{ic}}) \mathcal{L}_{\text{CTC}} + \lambda_{\text{ic}} \mathcal{L}_{\text{InterCTC}} + \lambda_{\text{ib}}\mathcal{L}_{\text{IB}}.
\label{eq:ctc_ae_loss}
\end{equation}

The IB loss is a variant of the InterCTC loss (Section \ref{sec:interctc}) with a significant difference, i.e., it minimizes the CTC objective using the bias phrase sequence instead of the actual target sequence. Additionally, the adapted formulation differs from the original in \cite{bias9,bias12} in two ways. Firstly, the former applies to the intermediate hidden states $\mathbf{h}_{(t,k)}^{'\text{AE}}$ of the audio encoder, giving equal weightage to potential contextual cues across different layers, thereby enhancing the biasing process. In contrast, the architecture in \cite{bias9, bias12} only considers the higher-level abstractions of the audio encoder at the $last$ ($N$) layer. Secondly, unlike in \cite{bias12}, where the authors do not modify the training targets for explicitly supervising the biasing task, our method provides explicit supervision. We force-align the bias phrase tokens with the frames to achieve a contextualized model. It ensures that all bias and non-bias frames contribute meaningfully to the loss.
\vspace*{-2mm}
\subsection{Contextual RNN-transducer}
We modify the original RNN-transducer \cite{rnnt1} network and leverage the contextual variant from \cite{bias4} for this study. Similar to \cite{bias4}, we incorporate two additional modules: a context encoder and a CB module as shown in Fig. \ref{fig:biasloss}(b). However, while \cite{bias4} only trained the CB module as a contextual adapter, we advocate for training the entire model from scratch. It allows all newly integrated components (including the proposed IB loss) to co-adapt, potentially leading to improved contextual performance. Following the processing of the CB module using Eq. \eqref{eq:crossattention}, we obtain the context-aware representations for the audio encoder $ \mathbf{h}_{(t,l)}^{'\text{AE}} $ and the predictor network $\mathbf{h}_{(u,l)}^{'\text{pred}}$ at $t$ and $u$, respectively. We integrate the CB module between both the encoder-joiner and predictor-joiner networks and update the joiner in Eq. \eqref{eq:jointnetwork}, enabling the model to predict the bias phrases within the input speech and is given by: 
\vspace*{-1mm}
\begin{equation}
\mathrm{P}(y_u|\mathbf{x}_{t},y_{(1:u-1,l)},\mathbf{h}_{(m,l)}^{\text{CE}}) = \mathrm{Softmax}(\mathrm{Joiner}(\mathbf{h}_{(t,l)}^{'\text{AE}},\mathbf{h}_{(u,l)}^{'\text{pred}})),
\label{eq:contextawarepredictions}
\vspace*{-1mm}
\end{equation}
We optimize the RNN-transducer loss using Eq. \eqref{eq:transducer_loss} given by:
\vspace*{-3mm}
\begin{equation}
\mathcal{L}_{\text{Tr}} = - \log \mathrm{P}_{\text{Tr}}(y_{(u,l)}|\mathbf{h}_{(t,l)}^{'\text{AE}}, \mathbf{h}^\text{CE}_{(m,l)}),
\label{eq:context_transducer_loss}
\end{equation}
Finally, the overall training objective is the weighted sum of Eq. \eqref{eq:ctc_ae_loss} and Eq. \eqref{eq:context_transducer_loss} with $\lambda_{\text{ae}}$ being the tunable hyper-parameter:
\vspace*{-2mm}
\begin{equation}
\mathcal{L} = \lambda_{\text{ae}}\mathcal{L}_{\text{AE}} + (1 - \lambda_{\text{ae}}) \mathcal{L}_{\text{Tr}}.
\label{eq:total}
\vspace*{-2mm}
\end{equation}
\subsection{RNN-transducer-driven joint decoding}
\vspace*{-1mm}
We adopt the RNN-transducer-driven joint decoding with CTC from \cite{4D} to mitigate the degradation of U-WER when the bias list size is $M=1000$ or large. In our proposed joint decoding we employ the contextual RNN-transducer decoder as the primary decoder and additionally account for CTC prefix scoring as proposed in \cite{ctc-attention1} to compute the CTC likelihoods. Subsequently, the hypotheses are scored by combining CTC and RNN-transducer decoders. The CTC score is added with a weight factor $\mu_{\text{ctc}}$ to the RNN-transducer having a weight factor of $\mu_{\text{tr}}$. Top $k_\text{beam}$ hypotheses are retained for the next time frame based on the obtained joint score with the main beam size.
\begin{table*}[t]
\caption{LS-960h test sets results following the test time biasing list selection in Section \ref{sec:experiments} with different sizes $M$. Reported metrics are in the following format: WER (U-WER/B-WER). \textbf{Bold:} the proposed method outperforms the baselines.  \textbf{\underline{Underlined:}} the best result.}
\vspace*{-7.0mm}
\label{maintable}
\begin{center}
\resizebox {0.95\linewidth} {!} {
\begin{tabular}{@{}l|l|cccc|cccc}
\toprule
\multirow{2}{*}{ID} & \multirow{2}{*}{Model Name} & \multicolumn{4}{c|}{test-clean} &\multicolumn{4}{c}{test-other} \\
      \cmidrule(lr){3-6}
      \cmidrule(lr){7-10}
  & & $M$=0 (no-bias) & $M$=100 & $M$=500 & $M$=1000 & $M$=0 (no-bias) & $M$=100 & $M$=500 & $M$=1000  \\
\midrule
\midrule
\multirow{2}{*}{\texttt{A1}} & InterCTC-based RNN-T  \cite{interctc4,bias4} & 3.68 & 3.68 & 3.68 & 3.68 & 8.56 & 8.56 & 8.56 & 8.56 \\
& (non-contextual baseline) & (2.18/15.83) & (2.18/15.83) & (2.18/15.83) & (2.18/15.83) & (5.88/32.11) & (5.88/32.11) & (5.88/32.11) & (5.88/32.11)\\
\midrule
\multirow{2}{*}{\texttt{A2}} & CPPNet-based RNN-T\footref{fn:reproduced}  \cite{bias9} & 3.93 & 3.28 & 3.52 & 3.67 & 9.12 & 7.97 & 8.30 & 8.41 \\
& (contextual baseline) & (2.52/15.36) & (2.28/11.42) & (2.20/14.16) & (2.64/15.13) & (6.66/30.82) & (6.05/24.79) & (5.96/28.86) & (5.96/29.98) \\
\midrule
\multirow{2}{*}{\texttt{B1}} & Intermediate Biasing (proposed) & \textbf{3.07} & \textbf{2.47} & \textbf{2.88} & \textbf{3.04} & \textbf{7.59} & \textbf{6.66} & \textbf{7.30} & \textbf{7.55}  \\
& ($\mathcal{K}$ = 2,4,6,8,10)  & (\textbf{1.84}/\textbf{13.07}) & (\textbf{1.69}/\textbf{8.85}) & (\textbf{1.80}/\textbf{11.63}) & (\textbf{1.88}/\textbf{12.58}) & (\textbf{5.29}/\textbf{27.81})  &(\textbf{4.99}/\textbf{21.36}) & (\textbf{5.18}/\underline{\textbf{26.00}}) & (\textbf{5.26}/\underline{\textbf{27.72}}) \\
\midrule
\multirow{2}{*}{\texttt{B2}} &  \texttt{B1} + joint decoding & \underline{\textbf{2.85}} & \underline{\textbf{2.31}} & \underline{\textbf{2.80}} & \underline{\textbf{2.84}} & \underline{\textbf{7.18}} & \underline{\textbf{6.40}} & \underline{\textbf{7.10}} & \underline{\textbf{7.34}} \\
& (proposed)  & (\underline{\textbf{1.60}}/\underline{\textbf{12.96}}) & (\underline{\textbf{1.53}}/\underline{\textbf{8.60}}) & (\underline{\textbf{1.62}}/\underline{\textbf{11.23}}) & (\underline{\textbf{1.66}}/\underline{\textbf{12.44}}) & (\underline{\textbf{4.80}}/\textbf{28.09}) & (\underline{\textbf{4.72}}/\underline{\textbf{21.18}}) & (\underline{\textbf{4.91}}/\textbf{26.30}) & (\underline{\textbf{5.00}}/\textbf{27.93}) \\
\bottomrule
\end{tabular}
}
\end{center}
\vspace*{-10mm}
\end{table*}
\vspace{-2mm}
\section{Experiments}
\label{sec:experiments}
\noindent \textbf{Dataset and evaluation metrics.} We evaluate our method on the LibriSpeech corpus (LS-960h) \cite{librispeech} (960 hours of English speech), a common speech recognition benchmark. We dynamically generate a biasing list during training, including randomly selected bias phrases (Section \ref{sec:context-encoder}) shared across utterances within the batch. This strategy promotes robustness to out-of-vocabulary (OOV) words. At testing, biasing lists are composed of rare words from the target utterance and distractors from \cite{shallow6}. We evaluate contextual biasing performance using WER, U-WER, and B-WER criterion, as established in \cite{shallow6}. Our goal is to minimize WER and B-WER without substantially increasing U-WER, even with a growing number of distractors.

\noindent \textbf{Experimental setup.}
We develop baseline and proposed model architectures using the ESPnet2 toolkit \cite{espnet} to present the proposed method's adaptability to InterCTC-based \cite{interctc4} contextual RNN-transducer \cite{bias4} models. The input features consists of 80-dimensional Mel filterbanks (window size 512 samples, hop length 160 samples), with SpecAugment applied for data augmentation. The audio encoder comprises two convolutional layers (stride two), a 256-dimensional linear projection, followed by 12 conformer layers (1024 linear units) with layer normalization.  Attention layers within the audio encoder utilized four MHA modules (dimension 256). The prediction network contains a single LSTM layer (256 hidden units), while the joint network has one layer (size 320).  The context encoder includes an embedding layer (size 256), a two-layer BiLSTM (size 256), and a linear projection layer. The CB module employs MHA (embedding size 256, 4 attention heads) and two linear layers projecting input/output to match embedding size and subsequent layer (size 256). Optimal training weights are determined empirically as  $\lambda_{\text{ae}}$ = 0.3, $\lambda_{\text{ic}}$ = 0.66, and $\lambda_{\text{ib}}$ = 0.03. During training, we uniformly extract 0 to 2 bias phrases for each utterance ($M_{\text{utt}}$) of $L_{\text{max}}=10$ token lengths per batch resulting in a total of $M$ bias phrases ($M_{\text{utt}} \times n_{\text{batch}}$). The InterCTC model is trained for 150 epochs while the RNN-transducer model for 70 epochs (limited by loss divergence), both using a 0.0015 learning rate, 25000 warmup steps, and the Adam optimizer.
\vspace{-3mm}
\section{Results}
\vspace{-1mm}
We first study the effect of the IB loss on the audio encoder and report our findings in Table \ref{ctc-table}. The IB loss, when adapted to an InterCTC-based model, achieves promising results on the LS-960h test sets. We have the following observations from these results: First, the random bias phrase extraction during training provides an augmentation effect, enabling the model to generate more generalizable representations. Second, the integration of the context encoder via the IB loss facilitates explicit alignment of general and context-specific bias embeddings due to the rules of CTC (Section \ref{sec:audioencoder}) and the advantages of the cross-attention mechanism (Section \ref{sec:CB}). This helps the encoder to understand the broader context and focus on the most relevant features.
\vspace*{-3.0mm}
\begin{table}[!h]
\renewcommand\thetable{1.1} 
\caption{Effect of the IB loss on audio encoder-only architecture. \textbf{Bold:} the proposed method outperforms the baselines.}
\vspace*{-7.0mm}
\label{ctc-table}
\begin{center}
\resizebox {0.95\linewidth} {!} {
\begin{tabular}{@{}lcccc}
    \toprule
    \multirow{2}{*}{Model} & \multicolumn{2}{c}{test-clean} &\multicolumn{2}{c}{test-other} \\
    \cmidrule(lr){2-3}
    \cmidrule(lr){4-5}
    & $M$=100 & $M$=1000 & $M$=100 & $M$=1000  \\
    \midrule
    InterCTC  \cite{interctc4} & 3.88 & 3.88 & 9.33 & 9.33 \\
    (non-contextual baseline) & (2.23/17.32) & (2.23/17.32) & (6.29/36.06) & (6.29/36.06)\\
    \midrule
    CTC-based CPPNet\footref{fn:reproduced} \cite{bias9} & 3.78 & 3.92 & 8.57 & 8.90 \\
    (contextual baseline) & (2.15/17.04) & (2.35/18.34) & (5.63/34.35) & (5.7/37.30) \\
    \midrule
    Intermediate Biasing (proposed) & \textbf{2.85} & \textbf{3.32} & \textbf{7.22} & \textbf{8.23} \\
    ($\mathcal{K}$ = 2,4,6,8,10)  & (\textbf{1.63}/\textbf{12.79}) & (\textbf{1.94}/\textbf{14.44}) & (\textbf{4.84}/\textbf{28.11})& (\textbf{5.49}/\textbf{32.24}) \\
    \bottomrule
\end{tabular}
}
\end{center}
\vspace*{-8mm}
\end{table}

In Table \ref{maintable}, we further compare our proposed method with the baseline non-contextual model ($\texttt{A1}$). Furthermore, for a fair comparison, we reproduced the contextual phrase prediction network (CPPNet) based on RNN-transducer ($\texttt{A2}$) \cite{bias9} following their experimental settings. However, it did not achieve a similar performance as described in their work.\footnote{Non-availability of source code and training pipeline limits our ability to identify potential shortcomings in our reproduced model.\label{fn:reproduced}} Table \ref{maintable} shows the proposed model ($\texttt{B1}$) reduces the WER from 3.68 to 2.47 (32.9\% relatively), U-WER from 2.18 to 1.69 (22.5\% relatively) and B-WER from 15.83 to 8.85 (44\% relatively) on test-clean set compared to $\texttt{A1}$ model when $M=100$. In addition, compared to the $\texttt{A2}$ baseline, our model also achieves substantial improvement in WER, U-WER, and B-WER. Although the proposed model becomes less effective as we increase the biasing list size $M=1000$, it still improves the performance from the contextual baseline \cite{bias9}.

\noindent\textbf{Effect of joint decoding.}
In Table \ref{maintable}, we investigate the impact of joint decoding on overall WER and U-WER when using a biasing list size of $M=1000$. We employ a beam size $k_\text{beam}$ of $10$ and decoder weights $\mu_{\text{ctc}}$=0.2 and $\mu_{\text{tr}}$=0.8 to perform joint decoding. Results show a substantial improvement in both metrics when employing RNN-transducer-driven joint decoding ($\texttt{B1} \text{vs} \texttt{B2}$). This indicates that joint decoding benefits from both the contextualized audio encoder and the predictor network. Importantly, joint decoding successfully improves U-WER from 1.88 to 1.66 (11.7\% relatively) when $M=1000$, while maintaining similar performance on B-WER. This finding suggests that joint decoding effectively mitigates the performance degradation in U-WER typically seen in contextual ASR systems with increasing the biasing list size.

\noindent\textbf{Effect of the IB loss in different layers.}
We investigate the effect of the IB loss in intermediate layers by employing it to three different combinations as shown in Table \ref{interctclayers}. We empirically find that employing the IB loss in alternative intermediate layer was most effective as it shows more improvement in WER, U-WER and B-WER. The observed performance gains suggest that the optimal placement of the IB loss is a critical factor in maximizing its effectiveness. This findings verify our main claim that model achieves and retains stronger contextualization ability if introduced early in the encoder layers with better alignment with the input audio.
\vspace*{-3.0mm}
\begin{table}[!h]
\caption{Effect of the IB loss on different intermediate encoder layers using LS-960h test sets. \textbf{Bold:} the proposed method outperforms the baseline. \textbf{\underline{Underlined:}} the best result. }
\vspace*{-6.0mm}
\label{interctclayers}
\begin{center}
\resizebox {0.95\linewidth} {!} {
\begin{tabular}{@{}lcccc}
    \toprule
    \multirow{2}{*}{Model} & \multicolumn{2}{c}{test-clean} &\multicolumn{2}{c}{test-other} \\
    \cmidrule(lr){2-3}
    \cmidrule(lr){4-5}
    & $M$=100 & $M$=1000 & $M$=100 & $M$=1000  \\
    \midrule
    InterCTC-based RNN-T & 3.68 & 3.68 & 8.56 & 8.56 \\
    (non-contextual baseline) \cite{interctc4,bias4} & (2.18/15.83) & (2.18/15.83) & (5.88/32.11) & (5.88/32.11)\\
    \midrule
    Intermediate Biasing & \textbf{2.79} & \textbf{3.12} & \textbf{6.80} & \underline{\textbf{7.51}} \\
    ($\mathcal{K}$ = 6)  & (\textbf{1.86}/\textbf{10.31}) & (\textbf{1.95}/\textbf{12.65}) & (\textbf{5.00}/\textbf{22.61})& (\textbf{5.29}/\underline{\textbf{26.95}}) \\
    \midrule
    Intermediate Biasing & \textbf{2.65} & \textbf{3.14} & \underline{\textbf{6.58}} & \textbf{7.61} \\
    ($\mathcal{K}$ = 4,8)  & (\textbf{1.78}/\textbf{9.72}) & (\underline{\textbf{1.85}}/\textbf{13.62}) & (\underline{\textbf{4.82}}/\textbf{22.00})& (\underline{\textbf{5.20}}/\textbf{28.80}) \\
    \midrule
    Intermediate Biasing & \underline{\textbf{2.47}} & \underline{\textbf{3.04}} & \textbf{6.66} & \textbf{7.55} \\
    ($\mathcal{K}$ = 2,4,6,8,10)  & (\underline{\textbf{1.69}}/\underline{\textbf{8.85}}) & (\textbf{1.88}/\underline{\textbf{12.58}}) & (\textbf{4.99}/\underline{\textbf{21.36}})& (\textbf{5.26}/\textbf{27.72}) \\
    \bottomrule
\end{tabular}
}
\end{center}
\vspace*{-7mm}
\end{table}

\noindent\textbf{Validation on Japanese dataset}.
Table \ref{castable} presents results from our in-house dataset. We collected 93 hours of spoken Japanese across diverse scenarios (e.g., meetings and morning assemblies). We combined this proprietary dataset with 581 hours from the Corpus of Spontaneous Japanese \cite{csj} and 181 hours from the Advanced Telecommunications Research Institute International (ATR) speech database \cite{atr}. We developed the proposed model following the experimental setup outlined in Section \ref{sec:experiments}. Similar to Table \ref{maintable}, our method outperforms the non-contextual baseline using a biasing list of 203 phrases (including names and technical vocabulary). The model shows improvement in B-CER from 23.65 to 19.31 (18.4\% relatively), with slight decreases in CER and U-CER. Joint decoding further reduces B-CER substantially to 16.93 (28.4\% relatively) while also improving CER and U-CER performance. The results confirm the method's effectiveness across linguistically very different English and Japanese datasets.
\vspace*{-3.0mm}
\begin{table}[!h]
\caption{Results obtained on our in-house dataset}
\vspace*{-7mm}
\label{castable}
\begin{center}
\resizebox {0.95\linewidth} {!} {
\begin{tabular}{@{}lccc}
\toprule
Model & CER & U-CER & B-CER \\
\midrule
InterCTC-RNN-T (non-contextual baseline) \cite{interctc4,bias4} & 10.15 & 8.34 & 23.65 \\
Intermediate Biasing ($\mathcal{K}$ = 2,4,6,8,10) (proposed) & 10.22 & 9.00 & \textbf{19.31} \\
\quad + joint decoding & \textbf{\underline{9.28}} & \textbf{\underline{8.23}} & \textbf{\underline{16.93}} \\
\bottomrule
\end{tabular}
}
\end{center}
\vspace*{-5mm}
\end{table}
\vspace*{-6mm}
\section{Conclusion}
In this work, we present a contextualized E2E ASR framework by leveraging intermediate representations within the encoder, enabling it to co-adapt to any E2E ASR architecture. A series of experiments with public and industrial data shows the effectiveness and robustness of auxiliary IB loss in improving the ASR accuracy and contextualization ability of the model without using an external language model. Compared to the alternative formulation \cite{bias9,bias12}, the proposed method is integrated into the intermediate layers of the encoder and outperforms on WER, U-WER and B-WER. Furthermore, we analyze the performance with an increasing biasing list size and propose joint decoding to mitigate the degradation of U-WER when using a large biasing list.

\clearpage


\bibliographystyle{IEEEtran}
\bibliography{refs}

\end{document}